# Masses and Thermodynamics Properties of Heavy Mesons in the Non-Relativistic Quark Model Using Nikiforov-Uvarov Method

M. Abu-Shady[1], T. A. Abdel-Karim[1], and Sh. Y. Ezz-Alarab[1]

Department of Mathematics and Computer Sciences, Faculty of Science, Menoufia University, Egypt[1]

## Abstract

Thermodynamics properties of heavy mesons are calculated within the framework of the *N*-dimensional radial Schrödinger equation. The Cornell potential is extended by including the quadratic potential plus the inverse of quadratic potential. The energy eigenvalues and the corresponding wave functions are calculated in the *N*-dimensional space using the Nikiforov-Uvarov (NV) method. The obtained results are applied for calculating the mass of spectra of charmonium, bottomonium, $b\bar{c}$, and $c\bar{s}$ mesons. The thermodynamics properties of heavy quarkonia such as, the mean energy, specific heat, free energy, and entropy are calculated. The effect of temperature and the dimensionality number on heavy mesons masses and thermodynamics properties is investigated. The obtained results are improved in comparison with other theoretical approaches and in a good agreement with experimental data. We conclude that the present potential well describes thermodynamic properties in the three-dimensional space and also the higher dimensional space.

## 1. Introduction

Thermodynamics is the branch of physics concerned temperature and their relation to energy. This branch plays an important role in high energy physics [1]. According to statistical quantum chromodynamics (QCD), nuclear matter may undergo a color deconfined partonic phase, the quark-gluon plasma (QGP), at sufficiently high temperature and/or density. Over the past few decades, strenuous efforts have been made to devise clean and experimentally viable signals that can unambiguously identify the existence of QCD phase transition and trace out its signatures. Charmonium (a bound state of charm and anti-charm quarks) suppression had been predicted as a signature for the deconfinement transition [2, 3].

The Schrödinger equation (SE) plays an important role in describing many phenomena as in high energy physics. Thus, the solutions of the SE are important for calculating mass of quarkonia and thermodynamic properties. To obtain the exact and approximate solutions of SE, the various methods have been used for specific potentials such as Nikiforov–Uvarov method [ 4 ], supersymmetry quantum mechanics [5], asymptotic iteration method [6], Laplace integral transform [7]. Recently, N-dimensional Schrodinger equation has received focal attention in the literature. The higher dimension studies facilitate a general treatment of the problem in such a manner that one can obtain the required results in lower dimensions just dialing appropriate with N.
 [8-14].

The thermodynamic properties play an essential role in describing quark- gluon plasma [15], in which the thermodynamic properties of light quarks are calculated in comparison with the strange quark matter. In Ref. [16], the thermodynamic properties of the QGP are performed based on the constituent quasiparticle model of the quark-gluon plasma. In addition, the thermodynamic properties are investigated in the framework chiral quark models such as in Refs. [17-19] and also in molecular physics using the relativistic and non-relativistic models [20-23].

In the present paper, we aim to calculate the N-dimensional Schrödinger equation analytically by using NU method, and apply the present results to find the properties of quarkonium particles such as masses and thermodynamic properties which are not considered in other works.

The paper is organized as follows: In Sec.2, the NU method is briefly explained. In Sec.3, the energy eigenvalues and the corresponding wave functions are calculated in the *N*-dimensional space. In Sec.4, the results are discussed. In Sec.5, the summary and conclusion are presented.

## 2. Theoretical Description of the Nikiforov-Uvarov (NU) Method

The NU method is briefly given that is a suitable method to obtain the solution of the second-order differential equation which has the following form.

$$\Psi''(s) + \frac{\bar{\tau}(s)}{\sigma(s)} \Psi'(s) + \frac{\tilde{\sigma}(s)}{\sigma(s)^2} \Psi(s) = 0, \quad (1)$$

where $\sigma(s)$ and $\tilde{\sigma}(s)$ are the polynomial of maximum second degree and $\bar{\tau}(s)$ is a polynomial of maximum first degree with a coordinate transformation $S=S(r)$ to find the particular solution of equation (1) by separation of variables, if we use the following transformation

$$\Psi(s) = \Phi(s)\chi(s). \quad (2)$$

Equ. (1) can be written as in Ref. [23] as following

$$\sigma(s)\chi''(s) + \tau(s)\chi'(s) + \lambda\chi(s) = 0, \quad (3)$$

$$\pi(s) = \frac{\acute{\sigma}(s) - \bar{\tau}(s)}{2} \pm \sqrt{\left(\frac{\acute{\sigma}(s) - \bar{\tau}(s)}{2}\right)^2 - \tilde{\sigma}(s) + K\sigma(s)}, \quad (4)$$

$$\lambda = \lambda_n = -n\tau(\grave{s}) - \frac{n(n-1)}{2}\sigma''(s), \quad n=0,1,2,\ldots \quad (5)$$

$\chi(s) = \chi_n(s)$ is a polynomial of n degree which satisfies the hypergeometric equation which has the following form

$$\lambda = K + \acute{\pi}(s), \quad (6)$$

the $\pi(s)$ is a polynomial of first degree. The values of K in the square-root of Eq. (4) is possible to calculate if the expressions under the square root are square of expressions. This is possible if its discriminate is zero {for details, see Ref. [4]}.

## 3. The Schrödinger Equation and the Extended Cornell Potential

The SE for two particles interacting via symmetric potential in the N-dimensional space takes the form as in Ref. [24]

$$\left[\frac{d^2}{dr^2} + \frac{N-1}{r}\frac{d}{dr} - \frac{L(L+N-2)}{r^2} + 2\mu(E - V(r))\right]\Psi(r) = 0, \quad (7)$$

where L, N, and μ are the angular momentum quantum number, the dimensionality number, and the reduced mass for the quarkonium particle, respectively. Setting wave Function $\Psi(r) = r^{\frac{1-N}{2}}R(r)$, the following radial SE is obtained

$$\left[\frac{d^2}{dr^2} + 2\mu(E - V(r)) - \frac{(L+\frac{N-2}{2})^2 - \frac{1}{4}}{2\mu r^2}\right]R(r) = 0, \quad (8)$$

the extended Cornell potential is suggested as follows. V(r) takes the form.

$$V(r)= ar - \frac{b}{r} + cr^2 + \frac{d}{r^2}, \qquad (9)$$

where a, b, c, and d are arbitrary constants to be determined later. The first term is a linear term for confinement feature and coulomb's potential that describes the short distances between quarks. The two terms are called Cornell potential. In the present work, we extend the Cornell potential to include the quadratic potential and the inverse quadratic potential which play an important role in improving quarkonium properties such as in Refs. [25, 26].

By substituting Eq. (9) into Eq. (8), we obtain

$$\left[\frac{d^2}{dr^2} + 2\mu\left(E - ar + \frac{b}{r} - cr^2 - \frac{d}{r^2} - \frac{\left(l+\frac{N-2}{2}\right)^2 - \frac{1}{4}}{2\mu r^2}\right)\right] R(r) = 0, \qquad (10)$$

let us assume that $r = \frac{1}{x}$ and $r_0$ is a characteristic radius of the meson. Then the scheme is based on the expansion of $\frac{1}{x}$ in a power series around $r_0$ i.e. around $\delta = \frac{1}{r_0}$. Eq. (10) takes the following equation

$$\left[\frac{d^2}{dx^2} + \frac{2x}{x^2}\frac{d}{dx} + \frac{2\mu}{x^4}\left(-A + Bx - C_1 x^2\right)\right] R(x) = 0, \qquad (11)$$

where

$$A = -\mu\left(E - \frac{3a}{\delta} - \frac{6a}{\delta^2}\right), \qquad B = \mu\left(\frac{3a}{\delta^2} + \frac{8C}{\delta^3} + b\right),$$

and

$$C_1 = \mu\left(\frac{a}{\delta^3} + \frac{c}{\delta^4} - \frac{b}{\delta} + d + \frac{\left(l+\frac{N-2}{2}\right)^2 - \frac{1}{4}}{2\mu}\right),$$

therefore

$$\pi = \pm\sqrt{(K + 2C_1)x^2 - 2Bx + 2A}, \qquad (12)$$

the constant K is chosen such as the function under the square root has a double zero, i.e. its discriminant the constant K is chosen such as the function under the square root has a double zero, i.e. its discriminant for bound state solutions, we choose the positive sign in the above equation so that the derivative is given.

$$\grave{\tau} = 2 - \frac{2B}{\sqrt{2A}}, \qquad (13)$$

$$\lambda = \lambda_n = -n\grave{\tau}(s) - \frac{n(n-1)}{2}\sigma''(s), \qquad n = 0,1,2,\ldots \qquad (14)$$

Form Eq. (14), we obtain

$$\lambda_n = -n\left(2 - \frac{2B}{2\sqrt{A}}\right) - n(n-1), \tag{15}$$

by using Eq. (5), we obtain from Eq. (6); $\lambda = \lambda_n$. The energy eigenvalue in the *N*-dimensional space is given.

$$E_{nl} = \frac{3a}{\delta} + \frac{6c}{\delta^2} - \frac{2\mu \left(\frac{3a}{\delta^2} + \frac{8c}{\delta^3} + b\right)^2}{\left[(2n+1) + \sqrt{1 + \frac{8\mu a}{\delta^3} + 4\left(\left(L + \frac{N-2}{2}\right)^2 - \frac{1}{4}\right) - 8\mu d + \frac{24\mu c}{\delta^4}}\right]^2}, \tag{16}$$

By taking a = 0 and N = 3, we obtain the results in Ref. [27]. At d = 0, we obtain the results of Ref. [28].

The radial of the wave function of Eq. (10) takes the following form.

$$R_{nl} = C_{nl}\, r^{\frac{-B}{\sqrt{2A}} - 1}\, e^{\sqrt{2A}\, r}\, \left(-r^2 \frac{d}{dr}\right)^n \left(r^{-2n + \frac{-2B}{\sqrt{2A}}} e^{-2\sqrt{2A}\, r}\right), \tag{17}$$

*where* $C_n$ is the normalization constant that is determined by

$$\int |R_{n\,l}(r)|^2\, dr = 1. \tag{18}$$

## 3. Thermodynamics properties

We study thermodynamics properties of extended Cornell potential, the partition function is given $Z = \sum_{n=0}^{[\lambda]} e^{-\beta E}$, where $\beta = \frac{1}{KT}$, K is the Boltzmann constant. The principal quantum number n ranges from 0, 1, 2,…, [$\lambda$], $\lambda = \frac{1}{2}\left[\sqrt{\frac{A_2}{A_1}} - A_3\right]$ where $A_1, A_2, A_3$ are defined in Eq. (20). In the classical limit, at high temperature T for large [$\lambda$] the sum can be replaced by an integral and $[\eta]$ can be replaced by $[\eta] = \lambda$ as used in Refs. [20, 21].

### 3.1 Partition function

$$Z(\beta) = \int_0^\lambda e^{-\beta E_n}\, d\lambda,$$

by substituting Eq. (16), we obtain

$$Z(\beta) = \frac{1}{2} e^{-A_1 \beta}\left(-A_3 e^{\frac{A_2 \beta}{A_3^2}} + e^{\frac{A_2 \beta}{(A_3 + 2\lambda)^2}}(A_3 + 2\lambda) + \sqrt{A_2}\sqrt{\pi}\sqrt{\beta}\left(\text{Erfi}\left[\frac{\sqrt{A_2}\sqrt{\beta}}{A_3}\right] - \text{Erfi}\left[\frac{\sqrt{A_2}\sqrt{\beta}}{A_3 + 2\lambda}\right]\right)\right), \tag{19}$$

where

$$A_1 = \left(\frac{3\,a}{\delta} + \frac{6\,a}{\delta^2}\right), \quad A_2 = 2\mu \left(\frac{3\,a}{\delta^2} + \frac{8\,C}{\delta^3} + b\right)^2,$$

$$A_3 = 1 + \sqrt{1 + \frac{8\,\mu a}{\delta^3} + 4\left(\left(L + \frac{N-2}{2}\right)^2 - \frac{1}{4}\right) - 8\mu d + \frac{24\,\mu c}{\delta^4}} \tag{20}$$

### 3.2 Mean energy U

$$U(\beta) = -\frac{d}{d\beta} Ln Z(\beta).$$

$$U(\beta) = -(2e^{A_1\beta}(\tfrac{1}{2}e^{-A_1\beta}(-\frac{A_2 e^{\frac{A_2\beta}{A_3^2}}}{A_3} + \frac{A_2 e^{\frac{A_2\beta}{(A_3+2\lambda)^2}}}{A_3+2\lambda} + \sqrt{A_2}\sqrt{\pi}\sqrt{\beta}\, D +$$

$$\frac{H}{2\sqrt{\beta}}) - \tfrac{1}{2}A_1 e^{-A_1\beta}(-A_3 e^{\frac{A_2\beta}{A_3^2}} + e^{\frac{A_2\beta}{(A_3+2\lambda)^2}}(A_3 + 2\lambda) +$$

$$\sqrt{\beta}\, H))) / (-A_3 e^{\frac{A_2\beta}{A_3^2}} + e^{\frac{A_2\beta}{(A_3+2\lambda)^2}}(A_3 + 2\lambda) + \sqrt{\beta}\, H)), \tag{21}$$

where

$$D = \frac{\sqrt{A_2}\, e^{\frac{A_2\beta}{A_3^2}}}{A_3\sqrt{\pi}\sqrt{\beta}} - \frac{\sqrt{A_2}\, e^{\frac{A_2\beta}{(A_3+2\lambda)^2}}}{\sqrt{\pi}\sqrt{\beta}(A_3+2\lambda)},$$

$$H = \mathrm{Erfi}\left[\frac{\sqrt{A_2}\sqrt{\beta}}{A_3}\right] - \mathrm{Erfi}\left[\frac{\sqrt{A_2}\sqrt{\beta}}{A_3+2\lambda}\right],$$

### (3.3) Specific heat C

$$C(\beta) = \frac{dU}{dT} = -K\beta^2 \frac{dU}{d\beta},$$

$$C(\beta) =$$

$$-K\beta^2((2e^{A_1\beta}(\tfrac{1}{2}e^{-A_1\beta}(-\frac{A_2 e^{\frac{A_2\beta}{A_3^2}}}{A_3} + \frac{A_2 e^{\frac{A_2\beta}{(A_3+2\lambda)^2}}}{A_3+2\lambda} + \sqrt{A_2}\sqrt{\pi}\sqrt{\beta}\, D +$$

$$\frac{\sqrt{A_2}\sqrt{\pi}\, H}{2\sqrt{\beta}}) - \tfrac{1}{2}A_1 e^{-A_1\beta}(-A_3 e^{\frac{A_2\beta}{A_3^2}} + e^{\frac{A_2\beta}{(A_3+2\lambda)^2}}(A_3 + 2\lambda) +$$

$$\sqrt{A_2}\sqrt{\pi}\sqrt{\beta}\, H)(-\frac{A_2 e^{\frac{A_2\beta}{A_3^2}}}{A_3} + \frac{A_2 e^{\frac{A_2\beta}{(A_3+2\lambda)^2}}}{A_3+2\lambda} + \sqrt{A_2}\sqrt{\pi}\sqrt{\beta}\, D + \frac{\sqrt{A_2}\sqrt{\pi}\, H}{2\sqrt{\beta}})) +$$

$$e^{\frac{A_2\beta}{(A_3+2\lambda)^2}}(A_3 + 2\lambda) + \sqrt{A_2}\sqrt{\pi}\sqrt{\beta}\, H)))//(-A_3 e^{\frac{A_2\beta}{A_3^2}} + e^{\frac{A_2\beta}{(A_3+2\lambda)^2}}(A_3 +$$

$$2\lambda) + \sqrt{A_2}\sqrt{\pi}\sqrt{\beta}\, H) \tag{22}$$

### (3.4) Free energy

F(β)= -KT LnZ(β),

F(β) =

$$-\frac{1}{\beta} \text{Log}\left[\frac{1}{2} e^{-A_1 \beta}\left(-A_3 e^{\frac{A_2 \beta}{A_3^2}} + e^{\frac{A_2 \beta}{(A_3+2\lambda)^2}}(A_3 + 2\lambda)\sqrt{A_2}\sqrt{\pi}\sqrt{\beta}\left(Erfi\left[\frac{\sqrt{A_2}\sqrt{\beta}}{A_3}\right] - Erfi\left[\frac{\sqrt{A_2}\sqrt{\beta}}{A_3+2\lambda}\right]\right)\right)\right].$$

(23)

### (3.5) Entropy

$S(\beta) = K \ln Z(\beta) + K T \frac{\partial}{\partial T} \ln Z(\beta).$

$S(\beta) =$

$$K\beta^2(-(2e^{A_1 \beta}(\frac{1}{2}e^{-A_1 \beta}(-\frac{A_2 e^{\frac{A_2\beta}{A_3^2}}}{A_3} + \frac{A_2 e^{\frac{A_2\beta}{(A_3+2\lambda)^2}}}{A_3+2\lambda} + \sqrt{A_2}\sqrt{\pi}\sqrt{\beta}\ D -$$

$$\frac{1}{2}A_1 e^{-A1\beta}(-A_3 e^{\frac{A_2\beta}{A_3^2}} + e^{\frac{A_2\beta}{(A_3+2\lambda)^2}}(A_3 + 2\lambda) + \sqrt{A_2}\sqrt{\pi}\sqrt{\beta}\ H\ )))/$$

$$((\beta(-A_3 e^{\frac{A_2\beta}{A_3^2}} + e^{\frac{A_2\beta}{(A_3+2\lambda)^2}}(A_3 + 2\lambda) + \sqrt{A_2}\sqrt{\pi}\sqrt{\beta}\ H\ )) +$$

$$\sqrt{A_2}\sqrt{\pi}\sqrt{\beta}\ H\ )) - \frac{1}{\beta}\text{Log}\left[\frac{1}{2}e^{-A_1\beta}\left(-A_3 e^{\frac{A_2\beta}{A_3^2}} + e^{\frac{A_2\beta}{(A_3+2\lambda)^2}}(A_3 + 2\lambda) + \sqrt{A_2}\sqrt{\pi}\sqrt{\beta}\ H\ \right)\right].$$

(24)

## 4. Results and Discussion
### [4.1] Quarkonium masses

In this subsection, we calculate spectra of the heavy quarkonium system such charmonium and bottomonium mesons that have the quark and antiquark flavor, the mass of quarkonium is calculated in 3-dimensional space (N=3 ). So we apply the following relation as in Refs. [23-25].

M=2m+$E_{nl}$, (25)

where, m is bare quark mass for quarkonium. By using Eq. (16), we can write Eq. (25) as follows

$$M = 2m + \frac{3a}{\delta} + \frac{6c}{\delta^2} - \frac{2\mu \left(\frac{3a}{\delta^2} + \frac{8c}{\delta^3} + b\right)^2}{\left[(2n+1) + \sqrt{1 + \frac{8\mu a}{\delta^3} + 4\left(\left(L + \frac{N-2}{2}\right)^2 - \frac{1}{4}\right) - 8\mu d + \frac{24 \mu c}{\delta^4}}\right]^2} \quad (26)$$

**Table (1)**: Mass spectra of charmonium (in GeV), ($m_c$=1.209 GeV, $a = 0.0.01\ GeV^2, b = 14.94, d = -15.04\ GeV^{-1}, c = 0.02\ GeV^3, \delta = 1.7\ GeV$.

| State | Present work | [8] | [27] | [29] | [24] | [32] | N=4 | Exp.[33] |
|---|---|---|---|---|---|---|---|---|
| 1s | 3.095 | 3.078 | 3.096 | 3.096 | 3.096 | 3.078 | 3.360 | 3.096 |
| 1p | 3.258 | 3.415 | 3.433 | 3.433 | 3.255 | 3.415 | 3.673 | - |
| 2s | 3.685 | 4.187 | 3.686 | 3.686 | 3.686 | 3.581 | 3.698 | 3.686 |
| 1D | 3.510 | 3.752 | 3.767 | 3.770 | 3.504 | 3.749 | 3.895 | - |
| 2p | 3.779 | 4.143 | 3.910 | 4.023 | 3.779 | 3.917 | 3.827 | 3.773 |
| 3s | 4.040 | 5.297 | 3.984 | 4.040 | 4.040 | 4.085 | 3.966 | 4.040 |
| 4s | 4.262 | 6.407 | 4.150 | 4.355 | 4.269 | 4.589 | 3.986 | 4.263 |
| 2D | 3.928 | - | - | 3.096 | - | 3.078 | 4.170 | 4.159 |

**In Table (1)**, we calculated energy eigenvalue from 1S to 2D. The charmonium mass is calculated by using Eq. (26). The free parameters of the present calculations are a, b, c and $\delta$ are fitted with experimental data. In addition, quark masses are obtained from Ref. [27]. In this Table (1), we note that calculated masses of charmonium are in good agreement with experimental data and are improved in comparison with Refs. [8, 24, 27, 29, 32], in which maximum error in comparison experimental data about 0.0555. The effect of dimensional number play an important role in the recent works [7-11]. The general form of higher dimensional gives more information about the system under study. In addition, we note that the charmonium mass increases with increasing dimensional number due to increasing binding energy. Therefore, when binding energy is larger than the constituents of charmonium which give us the limitation of non-relativistic models. The effect is also studied in Ref. [11].

**Table (2)**: Mass spectra of bottomonium (in GeV),( $m_b$=4.823 GeV, $a = 0.798 \text{ GeV}^2, b = 5.051, d = -3.854 \text{ GeV}^{-1}, c = 0.02 \text{ GeV}^3$, $\delta = 1.5$ GeV.

| State | Present work | [8] | [27] | [32] | [24] | [30] | N=4 | Exp. [31] |
|---|---|---|---|---|---|---|---|---|
| 1s | 9.460 | 9.510 | 9.460 | 9.460 | 9.460 | 9.510 | 9.610 | 9.460 |
| 1p | 9.609 | 9.862 | 9.840 | 9.811 | 9.619 | 9.862 | 10.022 | - |
| 2s | 10.022 | 10.627 | 10.023 | 10.023 | 10.023 | 10.038 | 10.072 | 10.023 |
| 1D | 9.846 | 10.214 | 10.140 | 10.161 | 9.864 | 10.214 | 10.205 | - |
| 2p | 10.109 | 10.944 | 10.160 | 10.374 | 10.114 | 10.396 | 10.269 | - |
| 3s | 10.360 | 11.726 | 10.280 | 10.355 | 10.355 | 10.566 | 10.306 | 10.355 |
| 4s | 10.580 | 12.834 | 10.420 | 10.655 | 10.567 | 11.094 | 10.344 | 10.580 |

**In Table (2)**, we note the present results for bottomonium are in agreement with experimental data, in which maximum error equals 0.0004828 and the present results are improved in comparison with Refs. [8, 24, 27, 32]. We note that the effect of dimensionality has the same effect as the charmonium.

**Table (3):**, mass spectra of b c̄ meson (in GeV)( $m_b = 4.823\ GeV$, $m_c = 1.209$ Gev, $a = 0.606\ GeV^2; b = 3.651, d = -2.199\ GeV^{-1}, c = 0.1\ GeV^3, \delta = 1.2\ GeV$ )

| State | Present work | [33] | [34] | [35] | N=4 | Exp. [31] |
|---|---|---|---|---|---|---|
| 1s | 6.277 | 6.349 | 6.264 | 6.270 | 6.355 | 6.277 |
| 1p | 7.042 | 6.715 | 6.700 | 6.699 | 6.883 | - |
| 2s | 7.383 | 6.821 | 6.856 | 6.835 | 6.878 | - |
| 2p | 6.663 | 7.102 | 7.108 | 7.091 | 7.161 | - |
| 3s | 7.206 | 7.175 | 7.244 | 7.193 | 8.035 | - |

**In Table (3),** we calculate mass spectra of meson bc̄ mesons where $2m = m_b + m_c$ in Eq. (26). We find that the 1S state closes with experimental data. The experimental data of the other states are not available. Hence, the theoretical predictions of the present method are displayed. We note that the present results of the bc̄ mass are in good agreement in comparison with Refs. [33-35].

**In Table** (4). Mass spectra of c s̄ meson in (GeV) ((m $_c$=1.628, m$_s$=0.419) GeV, a $= 0.48\ GeV^2$, b $= 3.795$, d $= -1.481\ GeV^{-1}$, c $= 0.1\ GeV^3$, δ $= 1.6\ GeV$).

| State | Present work | power | Screened | Phenomenological | N=4 | Exp. |
|---|---|---|---|---|---|---|
| 1s | 1.968 | 1.9724 | 1.9685 | 1.968 | 2.3001 | 1968.3[30] |
| 1p | 2.565 | 2.540 | 2.7485 | 2.566 | 2.742 | - |
| 2s | 2.709 | 2.6506 | 2.8385 | 2.815 | 2.797 | 2.709[36] |
| 3S | 2.932 | 2.9691 | 3.2537 | 3.280 | 2.967 | - |
| 1D | 2.857 | - | - | - | 3.934 | 2.859 [36] |

**In Table (4)**, we calculate mass spectra of cs̄ mesons where 2m=$m_c$+$m_s$ in Eq. (26). The 1S, 2S, 1D close with experimental data and are improved in comparison with power potential, screened potential, and phenomenological potential in Ref. [35]. The effect of dimensionality number has the same effect as in the charmonium and bottomonium.

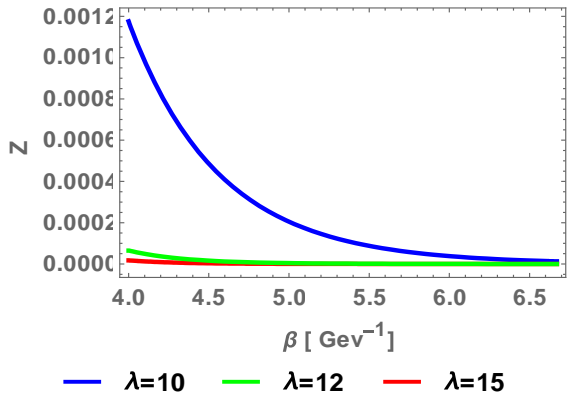

**Fig. 1:** Partition function is plotted as a function **of** β for different values of λ

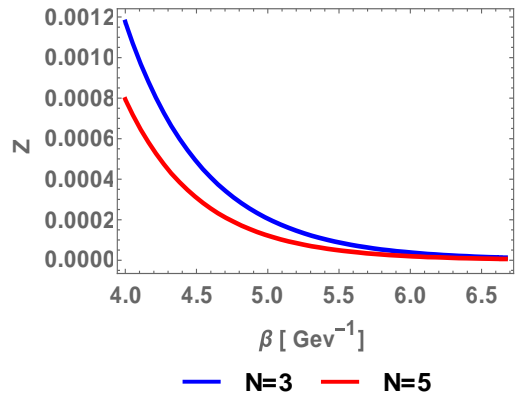

**Fig 2:** Partition function is plotted as a function of β for different values of N

**In Fig. (1),** we note that partition function Z decreases with increasing β. The range of β =4.0 to 6.5 GeV$^{-1}$ corresponding to T = 0.154 to 0.250 GeV which represents the range of temperature which charmonium melts to its constituents as charm quark. We note that the partition function decreases with increasing β and the dimensionality number N. In addition, we note the partition function is not sensitive at the largest values of β. The behavior of partition function will act on other observables that will be discussed.

In Ref. [20], It is observed that the partition function Z decreases monotonically with increasing β in which the author applied the deformed five-parameter exponential-type potential in the SE. In Ref. [21], the partition function decreases with increasing β. Therefore, the behavior of Z is in agreement with Refs. [20, 21]

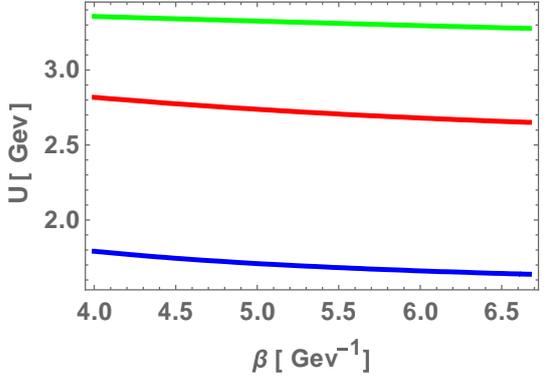 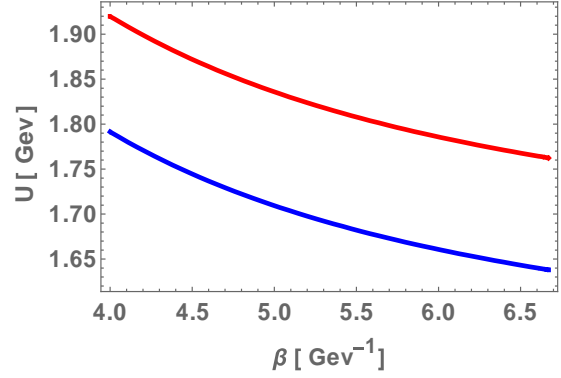

**Fig. 3:** Internal energy U against β for different values of λ    **Fig. 4:** Internal energy U against β for different values of N

**In Fig. (3),** we note that the U decreases with increasing of *β* and increases with increasing of *λ*. **In Fig. (4),** the values of U shifts to higher values by increasing dimensionality number. In Ref. [21], the authors found that the internal energy increases with increasing *λ* for HCL. In Ref. [20], the internal energy decreases with increasing *λ*. In Ref. [22], the authors calculated all thermodynamic properties of a neutral particle in a magnetic cosmic string background of using the non-relativistic Schrödinger–Pauli equation, in which they found that internal energy increases with increasing temperature and angular quantum number. We obtained the same conclusion in the present work for internal energy. Thus, the behavior of internal energy is agreement with recent works [20, 21, 22]. In Fig. (4), we note that the U increases with increasing dimensionality number. This effect is not considered in other works such as in Refs. [20, 21, 22].

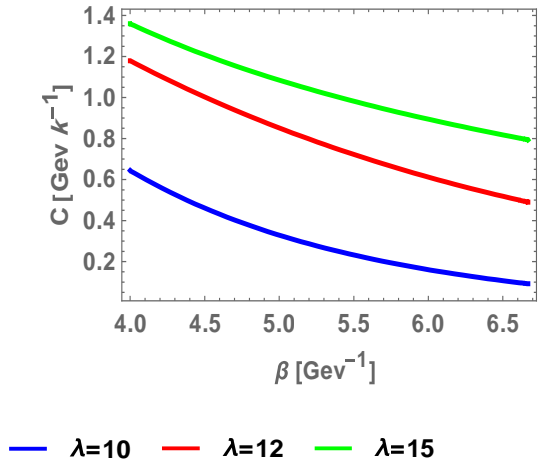 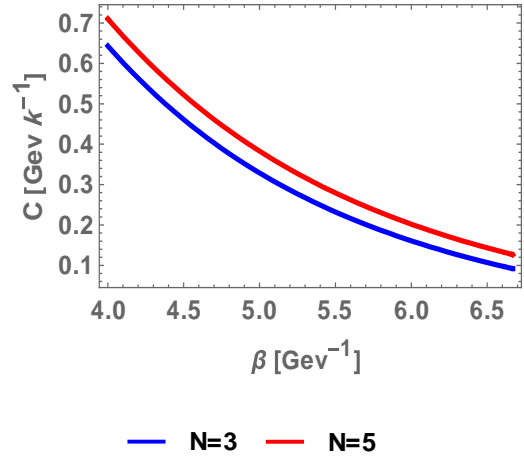

**Figure 5:** Specific heat C is plotted against $\beta$ for different values of $\lambda$

**Figure 6:** Specific heat C is plotted against $\beta$ for different values of N

**In Fig. (5),** we note that the specific heat (C) decreases with increasing of $\beta$ and it shifts to higher values by increasing parameter $\lambda$. In addition, the dimensionality number shifts the specific heat to higher values as **in Fig. (6)**. The behavior of the specific head is studied in Refs. [20, 21, 22] using the Schrödinger equation. We found the qualitative agreement with these works.

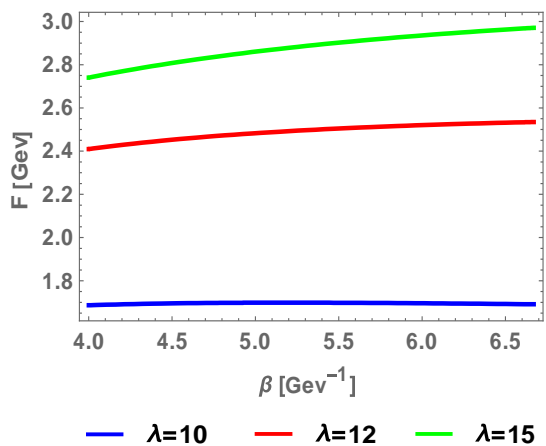 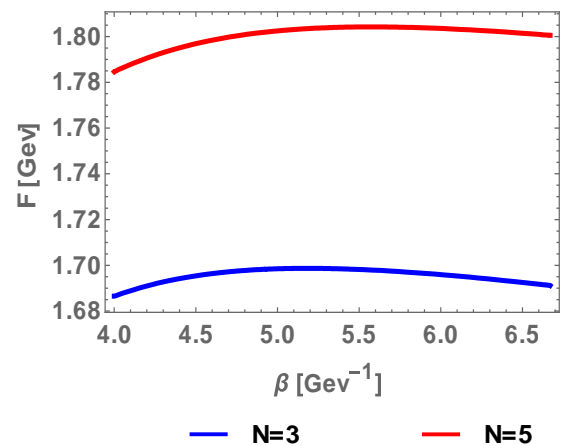

**Figure 7.** Free energy F is plotted against $\beta$ for different values of $\lambda$

**Figure 8.** Free energy F is plotted against $\beta$ for different values of N

**In Fig. (7),** we note that the free energy F increases with increasing $\beta$ and $\lambda$. Also, the free energy increases with increasing N, in **Fig. (8)**. In Ref. [15], the quark-gluon plasma is assumed to be composed of the light quarks only such as the up and down quarks, which interact weakly, and the gluons which are treated as they are free. They found the free energy decreases with increasing temperature. In Ref. [37], the authors studied free energy for strange quark matter and they found the conclusion of Ref. [15]. In Ref. [22], the authors calculated the free energy of a neutral particle and found the free energy decreases with increasing temperature. In present work, we note charm quark matter is qualitative agreement with Refs. [15, 22, 37].

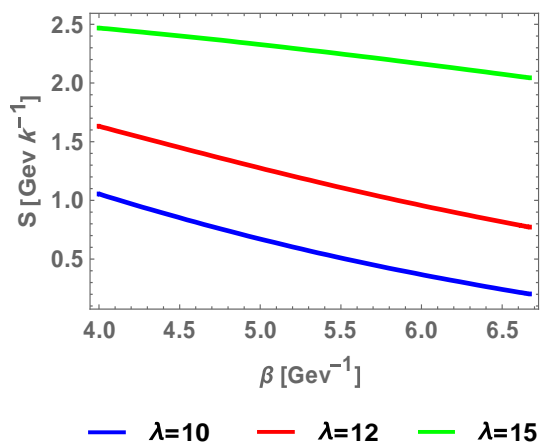
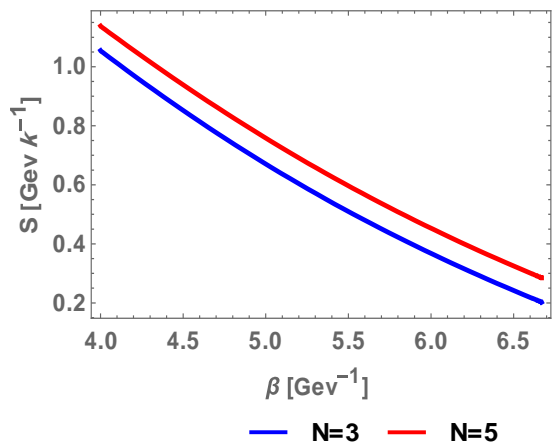

**Figure 9:** Entropy S is plotted against $\beta$ for different values of $\lambda$

**Figure 10**: Entropy S is plotted against $\beta$ for different values of N

**In Figs. (9, 10),** the entropy decreases with increasing $\beta$. In addition, the entropy shifts to higher values by increasing N and $\lambda$. In Refs. [15, 22, 37], the authors found the entropy increases with increasing temperature for light quark, strange quark, and natural particles. We have the same conclusion for charm quark.

## 5. Summary and Conclusion

In the present work, the energy eigenvalues and the wave functions are obtained in the *N*-dimensional space by solving the *N*-radial Schrödinger equation using the NU method. The Cornell potential is extended to include the quadratic potential and inverse quadratic potential which give a good description of heavy meson. We can obtain the energy eigenvalues and wave functions in lower dimension space. We applied the present results to calculate heavy-meson masses and thermodynamic properties. We conclude the following points:

**I**-At N = 3, the heavy meson masses are calculated. We found that the present results are in good agreement in comparison with experimental data. We found that the maximum errors are 0.0555 for the charmonium, 0.0004828 for the bottomonium, zero for the b c̄ meson and 0.000699 for the c s̄ meson. For N > 3, we found that the binding energy increases with increasing dimensional number space which leads to the limitation of non-relativistic quark models.

**II**- At N = 3, the thermodynamic properties are calculated such as internal energy, free energy, specific heat, and entropy. We found thermodynamic properties for charm quark plasma are qualitative agreement with Refs. [15, 22, 37]. In these works, thermodynamic properties the light quark, the strange quark, and natural particles are studied. Therefore, the thermodynamic properties for charmonium are not considered in many works such as [15, 22, 37]. For N > 3, the dimensionality number plays an important role in changing the behavior of thermodynamic properties for charmonium.

## 6. References

[1] W. Florkowski, Singapore: World scientific, 416 (2010).

[2] U. Kakade and B. K. Patra, Phys. Rev. C 92, 024901 (2015).

[3] T. Matsui and H. Satz, Phys. Lett. B 178, 416 (1986).

[4] A. F. Nikiforov and V. B. Uvarov, Spec. Func. Math. Phys. (Birkhauser, Basel 1988).

[5] F. Cooper, A. Khare, and U. Sukhatme, Phys. Rept. 251, 267 (1995)

[6] H. Ciftci, R. L. Hall, and N. Saad, J. Phys. A 36, 11807 (2003).

[7] T. Das, EJTP 13, 207 (2016).

[8] R. Kumar and F. Chand, Comm. Theor. Phys. 59, 528 (2013).

[9] M. Abu-Shady and E. M. Khokha, Adv. High . Ener. Phys, 7032041 12 (2018).

[10] M. Abu-Shady, T. A. Abdel-Karim, and E. M. Khokha, Adv. High Ener. Phys, 7356843 (2018).

[11] S. Roy and D.K. Choudhury, Cana. J. Phys. 94, 1282 (2016).

[12] Gen Gang, Chin. Phys. 14, 1075 (2005).

[13] G. R. Khan, Eur. Phys. J. D 53, 123 (2009).

[14] T. Das and A. Arda, Adv. High. Ener. Phys, 137038 (2015).

[15] M. Modarres and A. Mohamadnejad, Phys. Part. Nucl. Lett. 10, 99 (2013).

[16] V. S. Filinov, M. Bonitz, Y. B. Ivanov, M. Ilgenfritz, and V. E. Fortov, Contrib. Plasma Phys, 55, 203 (2015), hep-hp/1408571(2014).

[17] M. Abu-Shady, Inter. J. Theor. Phys. 54, 1530 (2015)

[18] M. Abu-Shady, Inter. J. Theor. Phys. 52, 1165 (2013).

[19] M. Schleif and R. Wunsch, Euro. Phys. J. A 1, 171 (1998).

[20] A. N. Ikot, B. C. Lutfuoglu, M. I. Ngwueke, M. E. Udoh, S. Zare, and H. Hassanabadi, Eur. Phys. J. Plus, 131, 419 (2016).

[21] W.A. Yah and K.J. Oyew, J. Asso. Arab. Univ. Bas. App. Scie. 21, 53 (2016).

[22] H. Hassanabadi and M. Hosseinpoura, Eur. Phys. J. C 76, 553 (2016).

[23] A. Boumali, EJT P. 12, 121,(2015).

[24] S. M. Kuchin and N. V. Maksimenko, Univ . J. Phys. Appl. **7**, 295 (2013).

[25] M. Abu-Shady, T. A. Abdel-Karim, and E. M. Khokha, SF. J. Quan Phys. 2, 1000017, ( 2018),

[26] P. Gupta and I. Mechrotra, J . Mod. Phys. 3, 1530 (2012).

[27] A. Al-Jamel and H. Widyan, Appl. Phys. Rese. 4, 94 (2013).

[28] M. Abu-shady, Inter. J. Appl. Math. Theo. Phys. 2, 16 (2016)

[29] N. V. Masksimenko and S. M. Kuchin, Russ. Phy. J. 54, 57 (2011).

[30] R. Kumar and F. Chand, Phys. Scr. 85, 055008 (2012).

[31] J. Beringer et al. [particle Data Group], Phys. Rev. D 86, 0100011, (2012).

[32] Z. Ghalenovi, A. A. Rajabi, S. Qin and H. Rischke, hep-ph/14034582 (2014).

[33] E. J. Eichten and C. Quigg, Phys. Rev. D 49, 5845 (1994).

[34] D. Ebert, R. N. Faustov, and V. O. Galkin, Phys. Rev. D 67, 014027 (2003).

[35] L. I. Abou-Salem, Inter. J. Mod. Phys. A 20, 4113 (2005).

[36] H. Manso and A. Gamal, Advan. High. Ener. Phys.

 7269657, 7, (2018).

[37] Modarres M., Gholizade H. Intern. J. Mod. Phys. E. 17, 1335, (2008)